\documentclass[english]{IEEEtran}
\usepackage{graphicx}
\usepackage{amssymb,amsbsy}
\usepackage{epstopdf}
\usepackage{amsmath,amsthm}
\usepackage{enumerate}
\usepackage{eufrak}
\usepackage{cite}
\usepackage{dsfont}
\usepackage{mathcomp}
\usepackage{supertabular}
\usepackage{makecell}
\usepackage{longtable}
\usepackage{stmaryrd}
\usepackage{url}
\usepackage{color}
\usepackage{rotating}
\usepackage{float}

\usepackage{array}
\usepackage{mathtools}
\usepackage{multicol}
\usepackage{amsmath}
\usepackage{algorithm}
\usepackage{algorithmic}
\usepackage{amssymb}
\usepackage{xcolor}

\usepackage[top=1.4cm, bottom=1.4cm, left=1.33cm, right=1.33cm]{geometry}

\usepackage[all]{xy}
\entrymodifiers={++[o][F-]}

\theoremstyle{definition}

\newtheorem{thm}{Theorem}
\newtheorem{cor}{Corollary}
\newtheorem{lem}{Lemma}

\newtheorem{defn}{Definition}
\newtheorem{cons}{Construction}
\newtheorem{exa}{Example}

\usepackage[T1]{fontenc}
\usepackage[latin9]{inputenc}
\usepackage{amsmath}
\usepackage{graphicx}
\usepackage{color}
\usepackage{multicol}



\usepackage{babel}



\usepackage{babel}

\begin{document}

\newcommand{\vA}{{\bf A}}
\newcommand{\vAtilde}{\widetilde{\bf A}}
\newcommand{\vB}{{\bf B}}
\newcommand{\vBtilde}{\widetilde{\bf B}}

\newcommand{\vC}{{\bf C}}
\newcommand{\vD}{{\bf D}}
\newcommand{\vH}{{\bf H}}
\newcommand{\vI}{{\bf I}}

\newcommand{\vY}{{\bf Y}}
\newcommand{\vZ}{{\bf Z}}

\newcommand{\vJ}{{\bf J}}

\newcommand{\vM}{{\bf M}}
\newcommand{\vN}{{\bf N}}
\newcommand{\vU}{{\bf U}}
\newcommand{\vV}{{\bf V}}
\newcommand{\vT}{{\bf T}}
\newcommand{\vR}{{\bf R}}
\newcommand{\vS}{{\bf S}}

\newcommand{\va}{{\bf a}}
\newcommand{\vb}{{\bf b}}
\newcommand{\vc}{{\bf c}}
\newcommand{\vd}{{\bf d}}

\newcommand{\ve}{{\bf e}}
\newcommand{\vh}{{\bf h}}
\newcommand{\vp}{{\bf p}}

\newcommand{\vu}{{\bf u}}
\newcommand{\vv}{{\bf v}}
\newcommand{\vw}{{\bf w}}
\newcommand{\vx}{{\bf x}}
\newcommand{\vhx}{{\widehat{\bf x}}}
\newcommand{\vtx}{{\widetilde{\bf x}}}
\newcommand{\vy}{{\bf y}}
\newcommand{\vz}{{\bf z}}

\newcommand{\vj}{{\bf j}}
\newcommand{\vzero}{{\bf 0}}
\newcommand{\vone}{{\bf 1}}
\newcommand{\vbeta}{{\boldsymbol \beta}}
\newcommand{\vchi}{{\boldsymbol \chi}}

\newcommand{\dA}{\mathtt A}
\newcommand{\dT}{\mathtt T}
\newcommand{\dC}{\mathtt C}
\newcommand{\dG}{\mathtt G}

\newcommand{\tA}{\textrm A}
\newcommand{\tB}{\textrm B}
\newcommand{\A}{\mathcal A}
\newcommand{\B}{\mathcal B}
\newcommand{\C}{\mathcal C}
\newcommand{\D}{\mathcal D}
\newcommand{\E}{\mathcal E}
\newcommand{\F}{\mathcal F}
\newcommand{\G}{\mathcal G}
\newcommand{\M}{\mathcal M}
\newcommand{\HH}{\mathcal H}
\newcommand{\PP}{\mathcal P}

\newcommand{\Q}{\mathcal Q}
\newcommand{\Qb}{\bar{\mathcal Q}}
\newcommand{\Db}{{\bar{\Delta}}}

\newcommand{\pQ}{{\bf p}\mathcal Q}
\newcommand{\pQb}{{\bf p}\bar{\mathcal Q}}

\newcommand{\R}{\mathcal R}
\newcommand{\SSS}{\mathcal S}
\newcommand{\U}{\mathcal U}
\newcommand{\V}{\mathcal V}
\newcommand{\Y}{\mathcal Y}
\newcommand{\Z}{\mathcal Z}

\newcommand{\Pg}{{{\mathcal P}_{\rm gram}}}
\newcommand{\Pgint}{{{\mathcal P}^\circ_{\rm gram}}}
\newcommand{\Pgrc}{{{\mathcal P}_{\rm GRC}}}
\newcommand{\Pgrcint}{{{\mathcal P}^\circ_{\rm GRC}}}
\newcommand{\Pint}{{{\mathcal P}^\circ}}
\newcommand{\Ag}{{\bf A}_{\rm gram}}

\newcommand{\CC}{\mathbb C} 
\newcommand{\RR}{\mathbb R}
\newcommand{\ZZ}{\mathbb Z}
\newcommand{\FF}{\mathbb F}
\newcommand{\KK}{\mathbb K}

\newcommand{\Fnd}{\FF_q^{n^{\otimes d}}}
\newcommand{\Knd}{\KK^{n^{\otimes d}}}

\newcommand{\ceiling}[1]{\left\lceil{#1}\right\rceil}
\newcommand{\floor}[1]{\left\lfloor{#1}\right\rfloor}
\newcommand{\bbracket}[1]{\left\llbracket{#1}\right\rrbracket}

\newcommand{\inprod}[1]{\left\langle{#1}\right \rangle}


\newcommand{\beas}{\begin{eqnarray*}} 
\newcommand{\eeas}{\end{eqnarray*}} 

\newcommand{\bm}[1]{{\mbox{\boldmath $#1$}}} 

\newcommand{\sizeof}[1]{\left\lvert{#1}\right\rvert}
\newcommand{\wt}{{\rm wt}} 
\newcommand{\supp}{{\rm supp}} 
\newcommand{\dg}{d_{\rm gram}} 
\newcommand{\da}{d_{\rm asym}} 
\newcommand{\dist}{{\rm dist}} 
\newcommand{\ssyn}{s_{\rm syn}}
\newcommand{\sseq}{s_{\rm seq}}
\newcommand{\nullplus}{{\rm Null}_{>\vzero}}

\newcommand{\tworow}[2]{\genfrac{}{}{0pt}{}{#1}{#2}}
\newcommand{\qbinom}[2]{\left[ {#1}\atop{#2}\right]_q}

\newcommand{\Lovasz}{Lov\'{a}sz }
\newcommand{\etal}{\emph{et al.}}

\newcommand{\todo}{{\color{red} (TODO) }}


\title{Weakly Mutually Uncorrelated Codes}

\author{
  \IEEEauthorblockN{
    S. M. Hossein~Tabatabaei Yazdi\IEEEauthorrefmark{1}, Han Mao Kiah\IEEEauthorrefmark{2}~and~
    Olgica~Milenkovic\IEEEauthorrefmark{1}}
  {\normalsize
    \begin{tabular}{ccc}
      \IEEEauthorrefmark{1}ECE Department, University of Illinois, Urbana-Champaign~~ &
      \IEEEauthorrefmark{2}SPMS, Nanyang Technological University, Singapore \\
    \end{tabular}}\vspace{-3ex}
    }
\maketitle
\begin{abstract}
We introduce the notion of weakly mutually uncorrelated (WMU) sequences, motivated by applications in DNA-based storage systems and synchronization protocols. WMU sequences are characterized by the property that no sufficiently long suffix of one sequence is the prefix of the same or another sequence. In addition, WMU sequences used in DNA-based storage systems are required to have balanced compositions of symbols and to be at large mutual Hamming distance from each other. We present a number of constructions for balanced, error-correcting WMU codes using Dyck paths, Knuth's balancing principle, prefix synchronized and cyclic codes. \end{abstract}
\IEEEpeerreviewmaketitle
\vspace{-0.21in}
\section{Introduction}
\vspace{-0.05in}
Mutually uncorrelated (MU) codes are a class of block codes in which no proper prefix of one codeword is a proper suffix of the same or another codeword.
MU codes were extensively studied in the coding theory and combinatorics literature under a variety of names.
Levenshtein introduced the codes in 1964 under the name `strongly regular codes'~ \cite{levenshtein1964decoding}, 
and suggested that the codes be used for synchronization.
Inspired by applications of distributed sequences in frame synchronization as described by van Wijngaarden and Willink in~\cite{de2000frame}, 
Baji\'c and Stojanovi\'c~\cite{bajic2004distributed} rediscovered mutually uncorrelated codes, and studied them under the name of 'cross-bifix-free' codes. Constructions and bounds on the size of MU codes were also reported in a number of recent contributions~\cite{bilotta2012new, blackburn2013non}. In particular, Blackburn ~\cite{blackburn2013non} analyzed these sequences under the name of `non-overlapping codes', and  provided a simple construction for a class of MU
codes with optimal cardinality. MU codes have also found applications in DNA storage~\cite{church2012next,goldman2013towards}: In this setting, Yazdi \etal~\cite{yazdi2015rewritable} developed a new, random-access and rewritable DNA-based storage architecture based on DNA sequences endowed with mutually uncorrelated address strings that allow selective access to encoded DNA blocks. The addressing scheme based on MU codes was augmented by specialized DNA codes in~\cite{kiah2015codes}.

Here, we generalize the family of MU codes by introducing weakly mutually uncorrelated (WMU) codes. WMU codes are block codes in which no ``long'' 
prefixes of one codeword are suffixes of the same or other codewords. WMU codes differ from MU codes in so far that they allow short prefixes of codewords to also appear as suffixes of codewords. This relaxation of prefix-suffix constraints was motivated in~\cite{yazdi2015rewritable} for the purpose of improving code rates while allowing for increased precision DNA fragment assembly and selective addressing. For more details regarding the utility of WMU codes in DNA storage, the interested readers are referred to the overview paper~\cite{yazdi2015dna}.

We are concerned with determining bounds on the size of WMU codes and efficient WMU code constructions. We consider both binary and quaternary WMU codes, the later class adapted for encoding over the four letters DNA alphabet $\{{\tt A,T,C,G}\}$. Our contributions include bounds on the largest size of WMU codes, construction of WMU codes that achieve the derived upper bound as well as results on three important constrained versions of WMU codes:
balanced WMU codes, error-correcting WMU codes and balanced, error-correcting WMU codes.
A binary string is called balanced if half of its symbols are zero. On the other hand, a DNA string is termed balance if it has a $50\%$ GC content, representing the percentage of symbols that are either $\tt G$ or $\tt C$. Balanced DNA strands are more stable than DNA strands with lower or higher GC content and they have lower sequencing error-rates. At the same time, WMU codes at large Hamming distance limit the probability of erroneous codeword selection. 

The paper is organized as follows.
In Section~\ref{sec:basics} we review MU and introduce WMU codes, and derive bounds on the maximum size of the latter family of codes. In addition, we outline a construction that meets the upper bound. In Section~\ref{sec:wmuerror} we describe constructions for error-correcting WMU codes, while in Section~\ref{sec:wmubalanced} we discuss balanced WMU codes. Our main results are presented in Section~\ref{sec:all}, where we first propose to use cyclic codes to devise an efficient construction of WMU codes that are both balanced and have error correcting capabilities. We then proceed to improve the cyclic code construction in terms of coding rate through decoupled constrained and error-correcting coding for binary strings. In this setting, we use Knuth's balancing technique~\cite{knuth1986efficient} and DC-balanced codes~\cite{immink2004codes}. 

\vspace{-0.12in}
\section{MU and WMU Codes: Definitions, Bounds and Constructions} \label{sec:basics}
Throughout the paper we use the following notation: $\FF_q$ denotes a finite field of order $q \geq 2$. 
If not stated otherwise, we tacitly assume that $q=2$, and that the corresponding field equals $\FF_2 = \left\{ 0,1 \right\}$. 
We let $\va=\left(a_{1},\ldots,a_{n}\right) \in\FF_q^n $ stand for a word of length $n$ over $\FF_q$, and $\va^j _i=\left(a_{i},\ldots,a_{j}\right)$, $1 \leq i \leq j \leq n$, stand for a substring of $\va$ starting at position $i$ and ending at position $j$. Moreover, for two arbitrary words $\va \in\FF_q^n,\vb \in\FF_q^m$ we use $\va \vb$ to denote a word of length $n+m$ generated by appending $\vb$ to the right-hand side of $\va$.  
\vspace{-0.16in}
\subsection{MU Codes}
We say that $\va=\left(a_{1},\ldots,a_{n}\right) \in\FF_q^n $ is self uncorrelated if no proper prefix of $\va$ matches its suffix, i.e., $\left(a_{1},\ldots,a_{i}\right) \neq \left(a_{n-i+1},\ldots,a_{n}\right)$, for all $ 1\leq i < n$. One can extend this definition to mutually uncorrelated sequences as follows: two not necessarily distinct words $\va , \vb \in \FF_q^n$ are mutually uncorrelated if no proper prefix of $\va$ appears as a suffix of $\vb$ and vice versa. Furthermore, we say that $\C\subseteq\FF_q^n$ is a mutually uncorrelated (MU) code if any two not necessarily distinct elements in $\C$ are mutually uncorrelated.

The maximum cardinality of MU codes was determined up to a constant factor by Blackburn~\cite[Theorem 8]{blackburn2013non}. For completeness, we state this result below.

\begin{thm}
Let $A_{MU}(n,q)$ denote the maximum size of MU codes over $\FF_q^n$, for $n\geq 1$ and $q\geq 2$. Then there exist constants $0< C_{1}< C_{2}$ such that 
\label{thm:M1}
\[
C_{1}\frac{ q^n}{n} \leq A_{MU}(n,q) \le C_{2}\frac{ q^n}{n}.
\]
\end{thm}
To motivate our WMU code design methods, we next briefly outline two known and one new construction of MU codes.

\begin{cons}\label{cons:C1} (Prefix-Balanced MU Codes)
 Bilotta \etal{}~\cite{bilotta2012new} described a simple construction for MU codes based on well known combinatorial objects termed 
{\em Dyck words}. A Dyck word is a binary string composed of $n$ zeros and $n$ ones such that no prefix of the word has more zeros 
than ones. By definition, a Dyck word necessarily starts with a one and ends with a zero.
Consider a set $\D$ of Dyck words of length $2n$ and define the following set of words of length $2n+1$,
\[\C_D\triangleq\{1\va: \va\in\D\}.\] 
Bilotta \etal{} proved that $\C_D$ is a MU code. An important observation is that MU codes constructed using Dyck words are inherently balanced or near-balanced. To more rigorously describe this property of Dyck words, recall that a Dyck word has \emph{height} at most $D$ if for any prefix of the word, the difference between the number of ones and the number of zeros is at most $D$. Hence, the disbalance of any prefix of a Dyck word is at most $D$, and the disbalance of an MU codeword in $\C_D$ is one.
Let Dyck$(n,D)$ denote the number of Dyck words of length $2n$ and height at most $D$. For fixed values of $D$, de Bruijn \etal{} \cite{bruijn1972average} proved that 
\begin{equation}
{\rm Dyck}(n,D)\sim\frac{4^n}{D+1}\tan^2 \left(\frac{\pi}{D+1}\right) \cos^{2n} \left(\frac{\pi}{D+1} \right).
\end{equation}
Here, $f(n)\sim g(n)$ denotes $\lim_{m\to\infty} f(n)/g(n)=1$. Hence, Billota's construction produces balanced MU codes. In addition, the construction ensures that every prefix of a codeword is balanced as well. By mapping $0$ and $1$ to $\{\mathtt{A},\mathtt{T}\}$ and $\{\mathtt{C},\mathtt{G}\}$, respectively, we obtain a DNA MU code.
\end{cons}
\begin{cons}\label{cons:C2} (General MU Codes, Levenshtein \cite{levenshtein1964decoding} and Gilbert \cite{gilbert1960synchronization}). Let $\ell,n,$ $1\leq \ell \leq n-1$, be two integers and let $\C \subseteq \FF_q^n$ be the set of all words $\va=\left(a_{1},\ldots,a_{n}\right)$ such that
\begin{enumerate}
\item $\left(a_{1},\ldots,a_{\ell }\right)  = \left(0,\ldots,0\right)$ 
\item $a_{\ell +1},a_{n} \neq 0$
\item The sequence $\left(a_{\ell+2},\ldots,a_{n-1}\right)$ does not contain $\ell$ consecutive zeros as a subword.
\end{enumerate}
\end{cons}
Then, $\C$ is an MU code. Blackburn~\cite[Lemma 3]{blackburn2013non} showed that for $\ell = \log_q 2n$ this construction is optimal. His proof relied on the observation that the number of strings $\left( a_{\ell+2},\ldots,a_{n-1}\right)$ that do not contain $\ell$ consecutive zeros as a subword exceeds $\frac{ \left(q-1\right)^{2}\left(2q-1\right)}{4nq^4}q^n$, thereby establishing the lower bound of Theorem \ref{thm:M1}.
It is straightforward to modify the second proposed code construction so as to incorporate error-correcting properties in the underlying MU code. We outline our new code modification below.
\begin{cons}\label{cons:F1} (Error-Correcting MU Codes)
Fix $t$ and $\ell$ to be positive integers and consider a binary $[n_H,s,d]$ code $\C$
of length $n_H=t(\ell-1)$, dimension $s$ and Hamming distance $d$.
For each codeword $\vb \in \C$,
we map $\vb$ to a word of length $n=(t+1)\ell+1$ given by
\[\va=0^\ell 1\vb^{\ell -1}_{1}1\vb^{2(\ell -1)}_{\ell}1\cdots\vb^{t(\ell -1)}_{(t-1)(\ell -1)+1}1.\]
Furthermore, we define $\C_{\rm parse}\triangleq \{\va :\vb\in \C\}$.
\end{cons}
It is easy to verify that $|\C_{parse}| = |\C_H|$, and that the code $\C_{parse}$ has the same minimum Hamming distance as $\C_H$, i.e., $d(\C_{parse})=d(\C_H)$. As $n_H$ was chosen so that 
$\C_{parse} \subseteq \left\{ 0,1 \right\}^n$. 
In addition, the parsing code $\C_{parse}$ is an MU code, since it satisfies all the constraints required by Construction~\ref{cons:C2}. 
To determine the largest asymptotic size of a parsing code, we briefly recall the Gilbery-Varshamov bound.
\begin{thm}\label{thm:GV}
(Asymptotic Gilbert-Varshamov bound \cite{gilbert1952comparison,varshamov1957estimate})
For any two positive integers $n$ and $d \leq \frac{n}{2},$ there exists a block code $\C \subseteq \left \{ 0,1 \right \}^n$ of minimum Hamming distance $d$ with normalized rate  
\begin{equation*}
R(\C) \geq 1 - h\left(\frac{d}{n}\right) -o(1),
\end{equation*}
where $h(\cdot)$ is an entropy function, i.e., $h(x)=x \log_2 \frac{1}{x} + (1-x) \log_2 \frac{1}{1-x}$, for $0 \leq x \leq 1$.
\end{thm}

\begin{cor}\label{cor:C2}
For a fixed value of $n$, $n_H$ is maximized in the aforementioned construction by choosing 
$\ell ^\ast= \sqrt{n-2}$; in this case, $n^{\ast}_{H} = (\sqrt{n-2}-1)^2 = n-2\sqrt{n-2}-1$.
By applying the GV result from Theorem \ref{thm:GV} and choosing $\C_H$ to be an $[n^{\ast}_{H} , s, d]$ block code, with $d \leq \frac{n^{\ast}_{H}}{2}$ and $s = n^{\ast}_{H} \, (1-h(\frac{d}{n^{\ast}_{H}}))$, we obtain an error-correcting MU code $\C_{parse}$ with parameters $[n,s,d]$.
\end{cor}   
\vspace{-0.14in}
\subsection{WMU Codes: Definitions, Bounds and Constructions}
The notion of mutual uncorrelatedness may be relaxed by requiring that 
only sufficiently long prefixes of one sequence do not match sufficiently long suffixes of other sequences. We next formally introduce codes with such defining properties.
\begin{defn}
Let $\C\subseteq\FF_q^n$ and $ 1\leq k \leq n$.
We say that $\C$ is a $k$-weakly mutually uncorrelated ($k$-WMU) code if 
no proper prefix of length $\ell$, for all $\ell \geq k$, of a codeword in $\C$ appears as a suffix of another codeword, including itself.
\end{defn}

\begin{thm}
\label{thm:M_2}Let $A_{WMU}\left(n,q,k \right)$ denote the maximum size of a $k$-WMU code over $\mathbb{F}_{q}^{n}$, for $n\geq 1$ and $q\geq 2$. Then, there exist constants 
$0<C_{3}<C_{4}$ such that 
\[
C_{3} \, \frac{q^{n}}{n-k+1}\leq A_{WMU}\left(n,q,k \right) \leq C_{4} \, \frac{q^{n}}{n-k+1}.
\]
\end{thm}
\begin{IEEEproof}
\label{proof_M_2_bounds}To prove the upper bound, we use an approach first suggested by Blackburn in~\cite[Theorem 1]{blackburn2013non}. 
Assume
that $\C \subseteq \mathbb{F}_{q}^{n}$ is a $k$-WMU code.
Let $L=\left( n+1\right)\left(n-k+1\right)-1$, and consider the set $X$ of pairs $\left(\va ,i\right)$ where $\va \in\mathbb{F}_{q}^{L}$,
$i\in\left\{ 1,\ldots,L\right\} $, and where the cyclic subword of $\va$ of length $n$ starting at position $i$
belongs to $\C$. Note that our choice of the parameter $L$ is governed by the overlap length $k$.

Note that $\left|X\right|=L\left|\C\right|q^{L-n}$, since there are
$L$ possibilities for the index $i$, $\left|\C\right|$ possibilities for
the word starting at position $i$ of $\va$, and $q^{L-n}$ choices
for the remaining $L-n\geq 0$ symbols in $\va$. Moreover, if $\left(\va,i\right)\in X,$
then $\left(\va,j\right)\notin X$ for $j\in\left\{ i\pm1,\ldots,i\pm n-k\right\} _{\textrm{mod }L}$ due to the weak mutual uncorrelatedness property.
Hence, for a fixed word $\va\in\mathbb{F}_{q}^{L}$, there
are at most $\left\lfloor \frac{L}{n-k+1}\right\rfloor$ different
pairs $\left(\va,i_{1}\right),\ldots,\left(\va,i_{\left\lfloor \frac{L}{n-k+1}\right\rfloor}\right)\in X$. This implies that $\left|X\right|\leq \left\lfloor \frac{L}{n-k+1}\right\rfloor q^{L}$. Combining the two derived constraints on the size of $X$, we obtain
$$\left|X\right|=L\left|\C\right|q^{L-n}\leq\left\lfloor \frac{L}{n-k+1}\right\rfloor q^{L}.$$
Therefore, $ \left|\C\right| \leq \frac{q^{n}}{n-k+1}$.

To prove the lower bound, we introduce a simple WMU code construction, outlined in Construction~\ref{cons:C3}. 
\begin{cons}\label{cons:C3}
Let $k,n$ be two integers such that $1\leq k \leq n$. A $k$-WMU code $\C \in \FF_q^n$ may be generated through a concatenation
 $\C=\left\{ \va \vb \mid \va \in \C^{\prime}, \vb \in \C^{\prime \prime}\right\} $, where $\C^{\prime}\subseteq\FF_q^{k -1}$ is unconstrained, and $\C^{\prime \prime}\subseteq\FF_q^{n-k +1}$ is an MU code. It is easy to verify that $\C$ is an $k$-WMU code with $\left|\C^{\prime}\right| \, \left|\C^{\prime \prime}\right|$ codewords.
\end{cons}
Let  $\C^{\prime} = \mathbb{F}_{q}^{k -1}$ and let $\C^{\prime \prime}\subseteq\FF_q^{n-k +1}$ be the largest MU code of size $A_{MU}\left(n-k +1,q\right)$.
Then, $\left|\C\right| = q^{k -1} \, A_{MU}\left(n-k +1,q\right)$. The claimed lower bound now follows from the lower bound of Theorem~\ref{thm:M1}, establishing that $\left|\C\right| \geq C_{1}\frac{q^{n}}{n-k+1}$ 
\end{IEEEproof}
\vspace{-0.05in}
\section{Error-Correcting WMU Codes}\label{sec:wmuerror}

We now turn our attention to WMU code design problems of interest in DNA-based storage. The collection of results in this section pertains
to WMU code constructions with error-correcting functionalities.

Let us start by introducing a mapping $\Psi$ that allows the DNA code design problem to be reduced to a binary code construction. For any two binary
strings $\va=\left(a_{1},\ldots,a_{s}\right), \vb=\left(b_{1},\ldots,b_{s}\right)\in\left\{ 0,1\right\} ^{s}$, $\Psi\left(\va,\vb\right):\left\{ 0,1\right\} ^{s}\times\left\{ 0,1\right\} ^{s}\rightarrow\left\{ \mathtt{A},\mathtt{T},\mathtt{C},\mathtt{G}\right\} ^{s}$
is an encoding function that maps the pair $\va, \vb$ to a DNA string $\vc=\left(c_{1},\ldots,c_{s}\right)\in\left\{ \mathtt{A},\mathtt{T},\mathtt{C},\mathtt{G}\right\} ^{s}$, according to the following rules:
\begin{equation}
\textrm{for }1\leq i\leq s,\: \vc_{i}=\begin{cases}
\mathtt{A} & \textrm{if }\left(\va_{i}, \vb_{i}\right)=\left(0,0\right)\\
\mathtt{C} & \textrm{if }\left(\va_{i}, \vb_{i}\right)=\left(0,1\right)\\
\mathtt{T} & \textrm{if }\left(\va_{i}, \vb_{i}\right)=\left(1,0\right)\\
\mathtt{G} & \textrm{if }\left(\va_{i}, \vb_{i}\right)=\left(1,1\right)
\end{cases}\label{eq:mapping}
\end{equation}
Clearly, $\Psi$ is a bijection and $\Psi ( \va, \vb ) \Psi ( \vc, \vd ) =\Psi ( \va \vc, \vb \vd)$.
The next lemma lists a number of useful properties of $\Psi$.
\begin{lem}
\label{lem:DNA_mapping_properties}Suppose that $\C_{1},\C_{2}\subseteq\left\{ 0,1\right\} ^{s}$
are two binary block code of length $s$. Encode each pair $\left( \va, \vb \right)\in \C_{1}\times \C_{2}$
using the DNA block code $\C_{3}=\left\{ \Psi\left( \va , \vb \right)\mid \va \in \C_{1}, \vb\in \C_{2}\right\} $. Then: 
\begin{enumerate}
\item $\C_{3}$ is balanced if $\C_{2}$ is balanced.
\item $\C_{3}$ is a  $k$-WMU code if either $\C_{1}$ or $\C_{2}$ is a $k$-WMU code.
\item If $d_{1}$ and $d_{2}$ are the minimum Hamming distances of $\C_{1}$
and $\C_{2}$, respectively, then the minimum Hamming distance of $\C_{3}$
is at least $\min\left(d_{1},d_{2}\right)$.
\end{enumerate}
\end{lem}
\begin{IEEEproof}
\label{proof_DNA_mapping_properties}
\begin{enumerate}

\item Any $\vc\in \C_{3}$ may be written as $\vc=\Psi\left( \va, \vb\right),$ where $\va\in \C_{1}, \vb\in \C_{2}$. According to (\ref{eq:mapping}),
the number of $G,C$ symbols in $\vc$ equals the number of  ones in $\vb$. Since $\vb$ is balanced, exactly half of the symbols in $c$ are $G$s and $C$s. This implies that $\C_{3}$ has a $50\%$ $GC$ content.

\item We prove the result by contradiction. Suppose that $\C_{3}$ is not a $k$-WMU code while $\C_{1}$ is a $k$-WMU code.
Then, there exist $\vc, \vc' \in \C_{3}$ such that a proper prefix of length at least $k$
of $\vc$ appears as a suffix of $\vc^{\prime}$. Alternatively, there exist nonempty strings $\vp, \vc_{0}, \vc_{0}^{\prime}$ such that $\vc= \vp \vc_{0} , \vc^{\prime}= \vc_{0}^{\prime} \vp$ and the length of $\vp$ is at least $k$.
Next, we use the fact $\Psi$ is a bijection and find binary strings
$\va, \vb, \va_{0}, \vb_{0}$ such that 
$$\vp=\Psi\left( \va, \vb\right), \vc_{0}=\Psi\left( \va_{0}, \vb_{0}\right), \vc_{0}^{\prime}=\Psi\left( \va_{0}^{\prime}, \vb_{0}^{\prime}\right).$$
Therefore,
\begin{align*}
\vc=\vp \vc_{0}= \Psi\left( \va, \vb\right) \Psi\left( \va_{0}, \vb_{0}\right) = \Psi\left( \va \va_{0}, \vb \vb_{0} \right),\\
\vc'=\vc'_{0} \vp= \Psi\left( \va'_{0}, \vb'_{0}\right) \Psi\left( \va, \vb\right)  = \Psi\left(  \va'_{0} \va,  \vb'_{0} \vb \right),
\end{align*}
where $\va \va_{0}, \va'_{0} \va \in \C_{1}$. This implies that the string $\va$ of length at least $k$ appears both as a proper prefix and suffix of two not necessarily distinct elements of $\C_{1}$. This contradicts the assumption that $\C_{1}$ is a $k$-WMU code. 
It is easy to verify that the same argument may be used for the case that $\C_{2}$ is a $k$-WMU code. 

\item For any two distinct words $\vc,\vc^{\prime}\in \C_{3}$ there exist $\va,\va^{\prime}\in \C_{1}, \vb, \vb^{\prime}\in \C_{2}$
such that $\vc=\Psi\left( \va, \vb\right), \vc^{\prime}=\Psi\left( \va^{\prime}, \vb^{\prime}\right)$.
The Hamming distance between $\vc, \vc^{\prime}$ equals
\begin{align*}
\sum_{1\leq i\leq s}\mathds{1}\left(\vc_{i}\neq \vc_{i}^{\prime}\right) & =\sum_{1\leq i\leq s}\mathds{1}\left(\va_{i}\neq \va_{i}^{\prime}\vee \vb_{i}\neq \vb_{i}^{\prime}\right)\\
 & \geq\begin{cases}
d_{1} & \textrm{if } \va\neq \va^{\prime}\\
d_{2} & \textrm{if } \vb\neq \vb^{\prime}
\end{cases} \geq \min\left(d_{1},d_{2}\right).
\end{align*}
\end{enumerate}
This proves the claimed result.
\end{IEEEproof}
\begin{cons}\label{cons:H2} (Decoupled Binary Code Construction)
For given integers $n$ and $k\leq n$, let $m=n-k+1$. As before, let $\va$, $\vb$ and $\vc$ denote the binary component words used in the encoding. We construct $\C\in\left\{ \mathtt{A},\mathtt{T},\mathtt{C},\mathtt{G}\right\} ^{n}$ according to the following steps:
\begin{enumerate}
\item 
Encode $\va$ using a binary block code $\C_{1} \subseteq \left \{ 0,1 \right \} ^{k-1}$ of length $k-1$,
and minimum Hamming distance $d$. Let $\Phi_{1}$ denote the encoding function, so that $\Phi_{1}\left(\va\right)\in \C_{1}$. 
\item 
Invoke Construction~\ref{cons:F1} with $n=m$ to arrive at a binary
MU code $\C_{2} \subseteq \left \{ 0,1 \right \} ^{m}$ of length $m$,
and minimum Hamming distance $d$. Encode $\vb$ using $\C_{2}$. Let $\Phi_{2}$ denote the encoding
function, so that $\Phi_{2}\left(\vb\right)\in \C_{2}$.
\item 
Encode $\vc$ using a binary
block code $\C_{3} \subseteq \left \{ 0,1 \right \} ^{n}$ of length $n$ and minimum Hamming distance $d$. Let $\Phi_{3}$ denote the encoding function, so that $\Phi_{3}\left(\vc\right)\in \C_{3}$.
\end{enumerate}
The output of the encoder performing the three outlined steps equals $\Psi\left(\Phi_{1}\left(\va\right) \Phi_{2}\left(\vb\right) , \Phi_{3}\left(\vc\right) \right)$.
\end{cons}
Next, we argue that $\C$ is a WMU code with guaranteed minimum Hamming distance properties.
\begin{lem}
\label{lem:H2}Let $\C\in\left\{ \mathtt{A},\mathtt{T},\mathtt{C},\mathtt{G}\right\} ^{n}$
denote the code generated by Construction \ref{cons:H2}. Then:
\begin{enumerate}
\item $\C$ is $k$-WMU code.
\item The minimum Hamming distance of $\C$ is at least $d$.
\end{enumerate}
\end{lem}
\begin{exa}\label{cor:H2}
In Construction~\ref{cons:H2}, let $\C_{1}$ and $\C_{3}$ be $\left[k-1, s_1 ,d\right]$ and $\left[n,s_3,d\right]$ block codes, respectively, where $s_1 = (k-1) \, (1-h(\frac{d}{k-1})) , s_3 = n \, (1-h(\frac{d}{n}))$ and $d \leq \frac{k-1}{2}$ satisfy the Gilbert-Varshamov bound of Theorem \ref{thm:GV}. Construct an $\left[m,s_2,d\right]$ block code $\C_2$ by using Corollary~\ref{cor:C2}, with $m = n-k+1, m^{\ast}_{H} = m-2\sqrt{m-2}-1, s_2 = m^{\ast}_{H} \, (1-h(\frac{d}{m^{\ast}_{H}}))$ and $d \leq \frac{m^{\ast}_{H}}{2}$. 
For this choice of component codes, the cardinality of $\C$ equals
\begin{align*}
|\C|= & 2^{ s_1 +  s_2 + s_3}= 2^{(k-1) \, (1-h(\frac{d}{k-1})) + m^{\ast}_{H} \, (1-h(\frac{d}{m^{\ast}_{H}}))+ n \, (1-h(\frac{d}{n}))}\\
= & \frac{4^{ n-\sqrt{n-k-1}-\frac{1}{2}}}{2^{(k-1) \, h(\frac{d}{k-1}) + m^{\ast}_{H} \, h(\frac{d}{m^{\ast}_{H}})+ n \, h(\frac{d}{n})}}
\end{align*}
\end{exa}
\section{Balanced WMU Codes}\label{sec:wmubalanced}
We begin this section by reviewing a simple method for constructing balanced binary
words, introduced by Knuth~\cite{knuth1986efficient} in 1986. In this scheme,
an $n$-bit binary string $\left(a_{1},\ldots,a_{n}\right)$ is sent to an encoder that inverts the
first $b$ bits of the data word ($\left(a_{1},\ldots,a_{n}\right)+1^{b}0^{n-b}$). 
The value of $b$ is chosen so that the encoded word has an equal
number of zeros and ones. Knuth proved that it is always possible to find an index $b$ that ensures a
balanced output. The index $b$ is represented by a balanced binary word $\left(b_{1},\ldots,b_{p}\right)$
of length $p$. To create the final codeword, the encoder prepends $\left(b_{1},\ldots,b_{p}\right)$
to $\left(a_{1},\ldots,a_{n}\right)+1^{b}0^{n-b}$. The receiver can
easily decode the message by first extracting the index $b$ from the
first $p$ bits and then inverting the first $b$ bits of the length-$n$ sequence.

Let $A\left(n,d,w\right)$ denote the maximum cardinality of a binary constant weight-$w$ code of length $n$ and even minimum Hamming
distance $d$. Knuth~\cite{knuth1986efficient} proved that
\begin{align}
&A\left(n,2,\frac{n}{2}\right)=\binom{n}{\frac{n}{2}} \approx \frac{2^{n+1}}{\sqrt{2 \, \pi} \, n^{\frac{1}{2}}} \notag 
\end{align}
which is a simple consequence of Stirling's approximation formula $n!\approx\sqrt{2\pi n}n^{n}e^{-n}$. Furthermore, Graham~\etal~\cite{graham1980lower} derived several bounds for the more general function $A\left(n,d,w\right)$. 
An updated list on the exact values and bounds on $A(n,d,w)$ may be found at {\tt http://codes.se/bounds/}.
In our future analysis, we use the well known Johnson~\cite{johnson1962new} bound.
\begin{thm}\label{thm:JO} (Johnson Bound) For $n\to\infty$, one has
\[
\frac{2^{n+1}}{\sqrt{2 \, \pi} \, n^{\frac{d-1}{2}}}
\leq A\left(n,d,\frac{n}{2}\right) \leq  
\frac{2^\frac{n+1}{2} \, e^\frac{n}{2}}{\sqrt{2 \, \pi} \, n^{\frac{d-1}{2}}}.
\]
\end{thm}
\begin{cons}\label{cons:H1} (Balanced WMU Codes)
For given integers $n$ and $k\leq n$, let $m=n-k+1$. As before, let $\va$ and $\vb$ denote the binary words used in the quaternary mapping described before. Construct a code $\C\in\left\{ \mathtt{A},\mathtt{T},\mathtt{C},\mathtt{G}\right\} ^{n}$ as follows: 
\begin{enumerate}
\item Encode $\va$ using a $k$-WMU code $\C_{1 } \subseteq \left\{0,1 \right \}^n$ of length $n$. For example, one may use Construction \ref{cons:C3} to generate $\C_{1 }$. Let $\Phi_{1}$ denote the encoding function, so that $\Phi_{1}\left(\va \right)\in \C_{1}$.
\item Encode $\vb$ using a balanced code $\C_{2}\subseteq \left\{0,1 \right \}^n$ of length $n$ and size $A\left(n,2,\frac{n}{2}\right)$.
Let $\Phi_{2}$ denote the encoding function, so that $\Phi_{2}\left(\vc \right)\in \C_{2}$.
\end{enumerate}
The output of the encoder is $\Psi\left(\Phi_{1}\left(\va\right) , \Phi_{2}\left(\vb\right) \right)$.
\end{cons}
\begin{lem}
\label{lem:H1}Let $\C\in\left\{ \mathtt{A},\mathtt{T},\mathtt{C},\mathtt{G}\right\} ^{n}$
denote the code generated by Construction~\ref{cons:H1}. Then,
\begin{enumerate}
\item $\C$ is a $k$-WMU code.
\item $\C$ is balanced.
\end{enumerate}
\end{lem}
We discuss next the cardinality of the code $\C$ generated by Construction~\ref{cons:H1}.
According to Theorem~\ref{thm:M_2}, one has $|\C_{1}| = C_{3} \, \frac{2^{n}}{n-k+1}$ for some constant $C_3 > 0$. The result is constructive. In addition, $|\C_{2}| \approx \frac{2^{n+1}}{\sqrt{2 \, \pi} \, n^{\frac{1}{2}}}$.
Hence, the size of $\C$ is bounded from below by:
\begin{align*}
C_{3}\, \frac{4^{n+1}}{\sqrt{2 \, \pi} \, (n-k+1) \,n^{\frac{1}{2}} }. \notag
\end{align*}
Next, we slightly modify the aforementioned construction and combine it with the Prefix-Balanced Construction \ref{cons:C1} to obtain a near-balanced $k$-WMU code $\C\in\left\{ \mathtt{A},\mathtt{T},\mathtt{C},\mathtt{G}\right\} ^{n}$ with parameter $D$. 
For this purpose, we generate $\C$ according to the Balanced WMU Construction \ref{cons:H1}. We set $\C_2 = \left\{ 0,1 \right\}^n$ and construct $\C_1$ by concatenating $\C'_1 \subseteq \left \{ 0,1 \right \}^{k-1}$ and $\C''_1 \subseteq \left \{ 0,1 \right \}^{n-k+1}$. Here, $\C'_1$ is balanced and $\C''_1$ is a near-balanced WMU code with parameter $D$. It is easy to verify that $\C$ is a near-balanced $k$-WMU DNA code with parameter $D$ and cardinality
\begin{align*}
|\C| = & |\C'_1| \, |\C''_1| \, |\C_2| =A(k-1,2,\frac{k-1}{2}) \, {\rm Dyck}(\frac{n-k}{2},D) \, 2^n\\
 \sim & \frac{4^{n} \, \tan^2 \left(\frac{\pi}{D+1}\right) \, \cos^{n-k} \left(\frac{\pi}{D+1} \right)}{\sqrt{2 \, \pi} \, (D+1) \, (k-1)^{\frac{1}{2}}}.
\end{align*}

\section{Balanced and Error-Correcting WMU Codes}\label{sec:all}
In what follows, we describe the main results of this paper, pertaining to constructions of balanced, error-correcting WMUs. The first construction is conceptually simple and it lends itself to efficient encoding and decoding procedures.
The second construction outperforms the first construction in terms of codebook size, and it utilizes the binary encoding functions described in the previous sections.
 \vspace{-0.14in}
\subsection{A Construction Based on Cyclic Codes}
The next construction uses ideas similar to Tavares' synchronization technique~\cite{tavares1968study}.
We start with a simple lemma and a short justification for that.

\begin{lem}\label{lem:cyclic}
Let $\C$ be a cyclic code of dimension $k$. Then the run of zeros in any nonzero codeword is at most $k-1$.
\end{lem}
\begin{IEEEproof}
Assume that there exists a non-zero codeword $c(x)$, represented in polynomial form, with a run of zeroes of length $k$. 
Since the code is cyclic, one may write $c(x)=a(x)g(x)$, where $a(x)$ is the information sequence corresponding to $c(x)$ and $g(x)$ is the generator polynomial. Without loss of generality, one may assume that the zeros run appears in positions $0,\ldots,k-1$, so that $\sum_{i+j=s}\,a_i\,g_j=0$, for $s\in \{{0,\ldots,k-1\}}$. The solution of the previous system of equations is
$a_0=a_1=\ldots=a_{k-1}=0$, contradicting the assumption that $c(x)$ is non-zero.
\end{IEEEproof}
\begin{cons}\label{cons:HanMao1}
Let $\C$ be an $[n,k-1,d]$ cyclic code and let $\ve=(1,0,\ldots,0)$.
Then $\C+\ve$ is a $k$-WMU code with distance $d$. 
\end{cons}
\begin{IEEEproof}
Suppose that on the contrary the code is $\C$ is not WMU. 
Then there exists a proper prefix $\vp$ of length at least $k$ such that 
both $\vp\va$ and $\vb \vp$ belong to $\C+\ve$.
In other words, $(\vp \va)-\ve$ and $(\vb \vp)-\ve$ belong to $\C$. 
Consequently, $(\vp \vb)-\ve'$ belongs to $\C$, where $\ve'$ is a cyclic shift of $\ve$.
Hence, by linearity of $\C$, $\vz\triangleq \vzero (\va-\vb)+\ve'-\ve$ belongs to $\C$.
Now, observe that the first coordinate of $\vz$ is one, and hence nonzero. 
But $\vz$ has a run of zeros of length at least $k-1$, which is a contradiction.
Therefore, $\C+\ve$ is indeed a $k$-weakly mutually uncorrelated code.
Since  $\C+\ve$ is a coset of $\C$, the minimum Hamming distance property follows immediately.
\end{IEEEproof}
To use the above construction to obtain balanced DNA codewords, we map the elements in 
$\FF_4$ to $\{{\tt A,T,C,G}\}$ via
\[0\mapsto {\tt A},\ 1\mapsto {\tt C},\ \omega\mapsto {\tt T},\ \omega+1\mapsto {\tt G}.\ \]
Let $\va$ be a word of length $n$. Then it is straightforward to see that the word $(\va,\va+\vone)$ 
has balanced $GC$ content. This leads to the simple construction described next.
\begin{cor}
Let $\C$ be an $[n,k-1,d]$ cyclic code over $\FF_4$ that contains the all ones vector $\vone$.
Then 
\[\{(\vc+\ve,\vc+\vone+\ve): \vc\in\C\}\]
is a $GC$ balanced, $k$-WMU code with minimum Hamming distance $2d$.
\end{cor}
\vspace{-0.2in}
\begin{table*}[t]
\caption{Summary of the proposed constructions for $q=4$.}
  \centering
\vspace{-0.1in}
  \begin{tabular}{*{5}{c}}

    Code & $k$-WMU & $k$-WMU + Error-Correcting & $k$-WMU + Balanced & $k$-WMU + Error-Correcting + Balanced \\
    \hline
    Rate & 
$C_1 \, \frac{4^{n}}{n-k+1}$ & 
$ \frac{4^{ n-\sqrt{n-k-1}-\frac{1}{2}}}{2^{(k-1) \, h(\frac{d}{k-1}) + m^{\ast}_{H} \, h(\frac{d}{m^{\ast}_{H}})+ n \, h(\frac{d}{n})}}$ 
& $C_{3}\, \frac{4^{n+1}}{\sqrt{2 \, \pi} \, (n-k+1) \,n^{\frac{1}{2}} }$ &
 $\frac{4^{n-\sqrt{n-k-1}}}{ \sqrt{2 \, \pi} \, 2^{(k-1) \, h(\frac{d}{k-1}) + m^{\ast}_{H} \, h(\frac{d}{m^{\ast}_{H}})} \, n^{\frac{d-1}{2}}}$ \\
    \hline
    Construction & 
 Construction \ref{cons:C3}&
 Construction \ref{cons:H2}&
Construction \ref{cons:H1}&
 Construction \ref{cons:H3}\\
    \hline
    Note & 
 $C_1 = \frac{3}{2^6}$ &
$ m^{\ast}_{H} = n-k-2\sqrt{n-k-1}$ &
$C_3 = \frac{3}{2^6}$ &
$ m^{\ast}_{H} = n-k-2\sqrt{n-k-1}$ \\
\label{table:1}
  \end{tabular}
\vspace{-0.3in}
\end{table*}
\subsection{The Decoupled Binary Code Construction}

The next construction is a combination of the binary code Constructions in~\ref{cons:H2} and~\ref{cons:H1}.
\begin{cons}\label{cons:H3}
For given integers $n$ and $k\leq n$, let $m=n-k+1$ and let $\va$, $\vb$ and $\vc$ be the binary component words.
 Next, construct $\C\in\left\{ \mathtt{A},\mathtt{T},\mathtt{C},\mathtt{G}\right\} ^{n}$ by applying the following steps:
\begin{enumerate}

\item 
Encode $\va$ using a binary block code $\C_{1} \subseteq \left \{ 0,1 \right \} ^{k-1}$ of length $k-1$,
and minimum Hamming distance $d$. Let $\Phi_{1}$ denote the encoding function, so that $\Phi_{1}\left(\va\right)\in \C_{1}$. 
\item 
Invoke Construction~\ref{cons:F1} with $n=m$ to generate an
MU code $\C_{2} \subseteq \left \{ 0,1 \right \} ^{m}$ of length $m$ and minimum Hamming distance $d$. Encode $\vb$ using $\C_{2}$. Let $\Phi_{2}$ denote the encoding
function, so that $\Phi_{2}\left(\vb\right)\in \C_{2}$.
\item 
Generate a codeword $\vc$ from a balanced code $\C_{3}$ of length $n$, minimum Hamming distance $d$ and of size $A\left(n,d,\frac{n}{2}\right)$. Let $\Phi_{3}$ denote the underlying encoding function, so that $\Phi_{3}\left(\vc \right)\in \C_{3}$.
\end{enumerate}
The output of the encoder is $\Psi\left(\Phi_{1}\left(\va\right) \Phi_{2}\left(\vb\right) , \Phi_{3}\left(\vc\right) \right)$.
\end{cons}
The following result is a consequence of Lemmas~\ref{lem:H1},~\ref{lem:H2}.
\begin{lem}
\label{lem:H3}Let $\C\in\left\{ \mathtt{A},\mathtt{T},\mathtt{C},\mathtt{G}\right\} ^{n}$
denote the code generated by Construction \ref{cons:H3}. Then,
\begin{enumerate}
\item $\C$ is a $k$-WMU code.
\item $\C$ is balanced.
\item The minimum Hamming distance of $\C$ is at least $d$.
\end{enumerate}
\end{lem}
\begin{exa}\label{cor:H3}
Construct $\C_1$ and $\C_2$ according to Example \ref{cor:H2}. 
The size of the code $\C$ equals
\begin{align*}
|\C|= & |\C_{1}|\, |\C_{2}| \,|\C_{3}|= 2^{s_1 + s_2 } \,  A(n,d,\frac{n}{2})\\
=& 2^{(k-1) \, (1-h(\frac{d}{k-1})) + m^{\ast}_{H} \, (1-h(\frac{d}{m^{\ast}_{H}}))}  \,  A(n,d,\frac{n}{2})\\
\geq & \frac{4^{n-\sqrt{n-k-1}}}{ \sqrt{2 \, \pi} \, 2^{(k-1) \, h(\frac{d}{k-1}) + m^{\ast}_{H} \, h(\frac{d}{m^{\ast}_{H}})} \, n^{\frac{d-1}{2}}}.
\end{align*}
The last inequality follows from the lower bound of Theorem~\ref{thm:JO}.
\end{exa}
\vspace{-0.2in}

\subsection{Concatenated Construction}

For a given integer $s\ge 1$, suppose that $\C_{0}$ is a balanced error correcting $k$-WMU code over $\mathbb{F}_{q}^{s}$ with minimum Hamming distance $d$. The code $\C_{0}$ may be obtained by using one of the two methods described in this section. Our goal is to obtain a larger family of balanced error-correcting $k$-WMU codes $\C\subseteq\FF_q^n$ by concatenating words in $\C_{0}$, where $n=s \, m$, $m \geq 1$. 
\begin{cons}\label{cons:H4}
Select subsets 
$\C_{1},\ldots,\C_{m}\subseteq \C_{0}$ such that
\begin{align*}
 & \C_{1}\cap \C_{m}= \emptyset \\
\textrm{and } & (\C_{1}\cap \C_{m-1}=\emptyset) \textrm{ or } (\C_{2}\cap \C_{m}=\emptyset)\\
\vdots\\
\textrm{and } & ( \C_{1}\cap \C_{2}=\emptyset)\textrm{ or } \ldots\textrm{ or } (\C_{m-1}\cap \C_{m}=\emptyset)
\end{align*}
\end{cons}
Let $\C=\left\{ \va_{1}\ldots \va_{m} \mid \va_{i}\in \C_{i}\textrm{ for }1\leq i\leq m\right\}.$ We claim that $\C$ is a balanced error-correcting $k$-WMU code over $\mathbb{F}_{q}^{n}$.

To clarify the result, notice that each element in $\C$ is created by concatenating $m$ strings, where each string belongs to $\C_0 \subseteq \mathbb{F}_{q}^{s}$. In addition, the words in $\C$ inherit the distance and balanced properties of $\C_0$. Therefore, $\C$ is balanced and has minimum Hamming distance at least $d$.

Next, for any pair of not necessarily distinct $\va,\vb \in \C$ and for $k\leq l < n$,
we show that $\va_{1}^{l}$ and $\vb_{n-l+1}^{n}$ cannot be identical. This establishes that the constructed concatenated code is WMU. Let $l=is+j,$ where $i=\left\lfloor \frac{l}{s}\right\rfloor$ and $0\leq j<s$. We consider three different scenarios for the index $j$:
\begin{itemize}
\item $j=0$; In this case, $1\leq i<m$. Therefore, 
$(\C_{1}\cap \C_{m-i+1}=\emptyset) \textrm{ or } \ldots\textrm{ or } (\C_{i}\cap \C_{1}=\emptyset)$
implies that $\va_{1}^{l}\neq \vb_{n-l+1}^{n}$. 
\item $0<j< k$; Again, one can verify that $1\leq i<m$. It is easy
to show that $\va_{l-s+1}^{l-j}$ is a suffix of length $s-j$ of a word in $\C_{0}$ and 
$\vb_{n-s+1}^{n-j}$ is a prefix of length $s-j$
of an element in $\C_{0}$. Since $k<s-j<s,$ one has $\va_{l-s+1}^{l-j}\neq \vb_{n-s+1}^{n-j}$.
Hence, $\va_{1}^{l}\neq \vb_{n-l+1}^{n}$.
\item $k \leq j<s$; In this case, $\va_{l-j+1}^{l}$ is a proper prefix of length
$j$ of a word in $\C_{0},$ and $\vb_{n-j+1}^{n}$ is a proper suffix
of length $j$ of an element in $\C_{0}$. Since $k\leq j<s,$ one has
$\va_{l-j+1}^{l}\neq \vb_{n-j+1}^{n}$ and $\va_{1}^{l}\neq \vb_{n-l+1}^{n}$.
\end{itemize}
We summarize the results of our constructions of WMU codes in Table~\ref{table:1}.
\vspace{-0.18in}


\begin{thebibliography}{10}
\providecommand{\url}[1]{#1}
\csname url@samestyle\endcsname
\providecommand{\newblock}{\relax}
\providecommand{\bibinfo}[2]{#2}
\providecommand{\BIBentrySTDinterwordspacing}{\spaceskip=0pt\relax}
\providecommand{\BIBentryALTinterwordstretchfactor}{4}
\providecommand{\BIBentryALTinterwordspacing}{\spaceskip=\fontdimen2\font plus
\BIBentryALTinterwordstretchfactor\fontdimen3\font minus
  \fontdimen4\font\relax}
\providecommand{\BIBforeignlanguage}[2]{{%
\expandafter\ifx\csname l@#1\endcsname\relax
\typeout{** WARNING: IEEEtran.bst: No hyphenation pattern has been}%
\typeout{** loaded for the language `#1'. Using the pattern for}%
\typeout{** the default language instead.}%
\else
\language=\csname l@#1\endcsname
\fi
#2}}
\providecommand{\BIBdecl}{\relax}
\BIBdecl

\bibitem{levenshtein1964decoding}
V.~Levenshtein, ``Decoding automata, invariant with respect to the initial
  state,'' \emph{Problemy Kibernet}, vol.~12, pp. 125--136, 1964.

\bibitem{de2000frame}
A.~J. De~Lind Van~Wijngaarden and T.~J. Willink, ``Frame synchronization using
  distributed sequences,'' \emph{Communications, IEEE Transactions on},
  vol.~48, no.~12, pp. 2127--2138, 2000.

\bibitem{bajic2004distributed}
D.~Baji{\'c} and J.~Stojanovi{\'c}, ``Distributed sequences and search
  process,'' in \emph{Communications, 2004 IEEE International Conference on},
  vol.~1.\hskip 1em plus 0.5em minus 0.4em\relax IEEE, 2004, pp. 514--518.

\bibitem{bilotta2012new}
S.~Bilotta, E.~Pergola, and R.~Pinzani, ``A new approach to cross-bifix-free
  sets,'' \emph{IEEE Transactions on Information Theory}, vol.~6, no.~58, pp.
  4058--4063, 2012.

\bibitem{blackburn2013non}
S.~R. Blackburn, ``Non-overlapping codes,'' \emph{arXiv preprint
  arXiv:1303.1026}, 2013.

\bibitem{church2012next}
G.~M. Church, Y.~Gao, and S.~Kosuri, ``Next-generation digital information
  storage in dna,'' \emph{Science}, vol. 337, no. 6102, pp. 1628--1628, 2012.

\bibitem{goldman2013towards}
N.~Goldman, P.~Bertone, S.~Chen, C.~Dessimoz, E.~M. LeProust, B.~Sipos, and
  E.~Birney, ``Towards practical, high-capacity, low-maintenance information
  storage in synthesized dna,'' \emph{Nature}, 2013.

\bibitem{yazdi2015rewritable}
S.~Yazdi, Y.~Yuan, J.~Ma, H.~Zhao, and O.~Milenkovic, ``A rewritable,
  random-access dna-based storage system,'' \emph{Scientific Reports}, vol.~5,
  no. 14138, 2015.

\bibitem{kiah2015codes}
H.~M. Kiah, G.~J. Puleo, and O.~Milenkovic, ``Codes for dna sequence
  profiles,'' \emph{arXiv preprint arXiv:1502.00517}, 2015.

\bibitem{yazdi2015dna}
S.~Yazdi, H.~M. Kiah, E.~R. Garcia, J.~Ma, H.~Zhao, and O.~Milenkovic,
  ``Dna-based storage: Trends and methods,'' \emph{Molecular, Biological, and
  Multi-Scale Communications, IEEE Transactions on}, to appear.

\bibitem{knuth1986efficient}
D.~E. Knuth, ``Efficient balanced codes,'' \emph{Information Theory, IEEE
  Transactions on}, vol.~32, no.~1, pp. 51--53, 1986.

\bibitem{immink2004codes}
K.~A.~S. Immink, \emph{Codes for mass data storage systems}.\hskip 1em plus
  0.5em minus 0.4em\relax Shannon Foundation Publisher, 2004.

\bibitem{bruijn1972average}
N.~de~Bruijn, D.~Knuth, and S.~Rice, ``The average height of planted plane
  trees,'' \emph{Graph Theory and Computing/Ed. RC Read}, p.~15, 1972.

\bibitem{gilbert1960synchronization}
E.~Gilbert, ``Synchronization of binary messages,'' \emph{Information Theory,
  IRE Transactions on}, vol.~6, no.~4, pp. 470--477, 1960.

\bibitem{gilbert1952comparison}
E.~N. Gilbert, ``A comparison of signalling alphabets,'' \emph{Bell System
  Technical Journal}, vol.~31, no.~3, pp. 504--522, 1952.

\bibitem{varshamov1957estimate}
R.~Varshamov, ``Estimate of the number of signals in error correcting codes,''
  in \emph{Dokl. Akad. Nauk SSSR}, vol. 117, no.~5, 1957, pp. 739--741.

\bibitem{graham1980lower}
R.~L. Graham and N.~Sloane, ``Lower bounds for constant weight codes,''
  \emph{Information Theory, IEEE Transactions on}, vol.~26, no.~1, pp. 37--43,
  1980.

\bibitem{johnson1962new}
S.~M. Johnson, ``A new upper bound for error-correcting codes,''
  \emph{Information Theory, IRE Transactions on}, vol.~8, no.~3, pp. 203--207,
  1962.

\bibitem{tavares1968study}
S.~Tavares, ``A study of synchronization techniques for binary cyclic codes,''
  Ph.D. dissertation, Thesis (Ph. D.)--McGill University, 1968.

\end{thebibliography}
%
%
\end{document}